\documentclass[aps,prb,reprint]{revtex4-2}
\usepackage{graphicx} 
\usepackage{amsmath} 
\usepackage{mathtools} 
\usepackage{gensymb} 
\usepackage{bm} 
\usepackage{natbib}
\usepackage{natmove}
\usepackage[colorlinks=true, linkcolor=blue, citecolor=blue, urlcolor=blue, linktoc=page, bookmarks=false, pdfstartview={FitH}, pdfborder={0 0 0.0 [3 3]}]{hyperref} 
\usepackage{color} 
\usepackage{dcolumn} 
\newcolumntype{.}{D{.}{.}{2.1}}
\newcolumntype{-}{D{.}{.}{4.0}}
\usepackage{makecell}
\usepackage{multirow}
\usepackage{cleveref}
\usepackage{easyReview} 
\crefname{figure}{Fig.}{Figs}
\crefname{table}{Table}{Tables}
\crefname{equation}{Eq.}{Eqs.}
\crefname{section}{Sec.}{Secs.}
\renewcommand{\today}{\number\day \space \ifcase \month \or January\or February\or March\or April\or May\or June\or July\or August\or September\or October\or November\or December\fi \space \number\year} 
\def\m1r{\multicolumn{1}{r}}
\begin{document}
\title{Mapping Rashba and Dresselhaus spin-orbit interactions to inversion asymmetry in perovskite oxide heterostructures}
\author{Nirmal \surname{Ganguli}}
\email[Contact author: ]{NGanguli@iiserb.ac.in}
\author{Avishek \surname{Singh}}
\author{Vivek \surname{Kumar}}
\author{Jayita \surname{Chakraborty}}
\affiliation{Department of Physics, \href{https://ror.org/02rb21j89}{Indian Institute of Science Education and Research Bhopal}, Bhauri, Bhopal 462066, India}
\date{\today}
\begin{abstract}
Inversion asymmetry, combined with spin-orbit interaction, leads to Rashba or Dresselhaus effects or combinations that are promising for technologies based on antiferromagnetic spintronics. Since understanding the exact nature of spin-orbit interaction is crucial for developing a technology based on it, mapping the nature of inversion asymmetry with the type of spin-orbit interaction becomes the key. We simulate a perovskite oxide heterostructure LaAlO$_3|$SrIrO$_3|$SrTiO$_3$ preserving the inversion symmetry within density functional theory to demonstrate the relation between the nature of inversion asymmetry and the corresponding Rashba or Dresselhaus-type interaction. With progressive distortion in the heterostructure, we find how the structure inversion asymmetry sets in with distorted bond lengths and angles, leading to the Rashba effect in the system in the presence of a microscopic electric field. Further, the introduction of tilted IrO$_6$ octahedra leads to bulk inversion asymmetry, helping a combined Rashba-Dresselhaus interaction to set in. A comparison of the spin textures obtained from our DFT calculations and theoretical modeling helps us identify the exact nature of the interactions. In the context of perovskite heterostructures, our work demonstrates what type of distortions lead to bulk inversion asymmetry and what other distortions can lead only to structure inversion asymmetry by mapping different types of spin-orbit interactions to different distortions. Additionally, our work emphasizes the importance of spin texture in identifying the nature of spin-orbit interaction. It may serve as a guide to identifying different types of Rashba-like spin-orbit interactions.
\end{abstract}

\maketitle
\section{\label{sec:intro}Introduction}
Antiferromagnetic spintronics holds enormous promise as an upcoming quantum technology owing to ultrafast operation, nonvolatile memory, and extremely low power consumption in a proposed device \cite{BaltzRMP18, GomonayPSSRRL17, JungwirthNN16, SmejkalNP18}. An adjustable spin-orbit torque to manipulate a noncollinear antiferromagnetic texture, a requirement for device operation and memory, may be realized with the help of a Rashba-like spin-orbit interaction leading to a helical spin arrangement \cite{BaltzRMP18}. While structure inversion asymmetry (SIA) and bulk inversion asymmetry (BIA) are known to result in Rashba- and Dresselhaus-type interactions, respectively \cite{RashbaSPSS60, DresselhausPR55, WinklerSOC03}, finding the exact type and order of interaction and the underlying type of asymmetry causing the same often becomes a challenge, resulting in conjectures about the interaction instead of a confirmation. However, knowing the exact nature of the interaction is paramount for understanding the system and designing applications. For example, the spin-orbit torque and the transport properties, including anisotropic magnetoresistance and planar Hall effect, extensively depend on the type and the order of the interaction \cite{BoudjadaPRB19}, emphasizing the need for a mechanism that can unequivocally confirm the nature of the interaction.

Angle-resolved photoemission spectroscopy (ARPES) has been a characterizing tool for Rashba-like interactions for years. However, the technique probes only the quasiparticle energy in the momentum space, revealing limited information about the system's quantum state. Recently, an improvement came in the form of spin-resolved ARPES, allowing for probing the spin projections of some of the bands \cite{TakayamaBook17}. Nevertheless, extracting an exhaustive set of features essential for confirming the exact nature of interaction may be prohibitively difficult. Until recently, theoretical tools for understanding Rashba-like interactions were limited to analyzing a conventional band dispersion along certain high-symmetry lines and fitting the same to a symmetry-adapted $\vec{k} \cdot \vec{p}$ model. Besides a symmetry-adapted model better suiting only bulk systems, the model can reveal the projected spin texture only up to the accuracy of fitting the energy eigenvalues, thereby leaving the possibility of qualitative flaws. Projected spin textures directly obtained from density functional theory (DFT) calculations may serve as a confirmatory test to find the nature of Rashba-like interaction in a system \cite{KumarPRB22}.

The surfaces and interfaces of perovskite oxides with the general formula ABO$_3$ host Rashba-like interactions of various nature, order, and magnitude \cite{KumarPRB22, ChakrabortyPRB20, ShanavasPRL14, ZhongPRB13, KimPRB14, ZhouPRB15bb}. Additionally, owing to the two-dimensional conducting system at the interface and matured synthesis techniques, electronics using perovskite oxide heterostructures emerged as a promising research direction \cite{OhtomoN04, MannhartS10}. Interesting magnetic properties have also been observed at such interfaces, shaping them as potential candidates for technology \cite{BrinkmanNM07, GanguliPRL14, ChakrabortyPRB20}. A host of possible distortions in perovskite oxides make them an ideal subject for studying several interesting physical properties and their interrelation with distortion. For example, KTaO$_3$ is usually found in a perfect cubic structure, YMnO$_3$ shows Jahn-Teller (JT) distortion, SrTiO$_3$ exhibits tetragonal distortion, and CaTiO$_3$ hosts tilted TiO$_6$ octahedra. Additionally, inversion asymmetry arises in the structure due to the formation of interfaces or surfaces. Microscopic electric fields developed in the layered polar oxides also help manifest the Rashba interaction. Thus, perovskite oxide heterostructures offer a platform to systematically investigate and verify the specific types of asymmetry and their implications on Rashba-like spin-orbit interaction.

In view of a unique platform offered by perovskite oxide heterostructures for understanding Rashba-like interactions in connection with structural asymmetry, we thoroughly investigate various possible asymmetries in LaAlO$_3|$SrIrO$_3|$SrTiO$_3$ (LAO$|$SIO$|$STO) heterostructures \cite{ChakrabortyPRB20}, the corresponding changes in the electronic structure, magnetic properties, and Rashba-like effect using DFT calculations along with analytical and numerical models. Starting from a heterostructure comprising the components in an ideal cubic structure, hereafter referred to as the undistorted heterostructure, we simulate progressively distorted structures by accommodating JT-like tetragonal distortions, distorted bond angles, and tilted BO$_6$ octahedra. Besides carefully analyzing the electronic structure and magnetic properties, the three-dimensional (3D) band dispersions, isoenergetic contours, and the projected spin textures directly obtained from our DFT calculations unequivocally establish the nature of Rashba-like interactions and their connection with the specific type of asymmetry found in perovskite oxides. We organize the remainder of the article as follows: \cref{sec:method} details the methodology adopted for our calculations. We discuss the results of our calculations in \cref{sec:results}. The investigation is summarized and concluded in \cref{sec:conc}. Finally, various combinations of Rashba-Dresselhaus models of different orders are discussed in the \cref{sec:RDmodels}illustrating the corresponding energy eigenvalues and projected spin textures that facilitate our analysis in \cref{sec:results}.

\section{\label{sec:method}Method}
\begin{figure}
	\includegraphics[scale = 0.33]{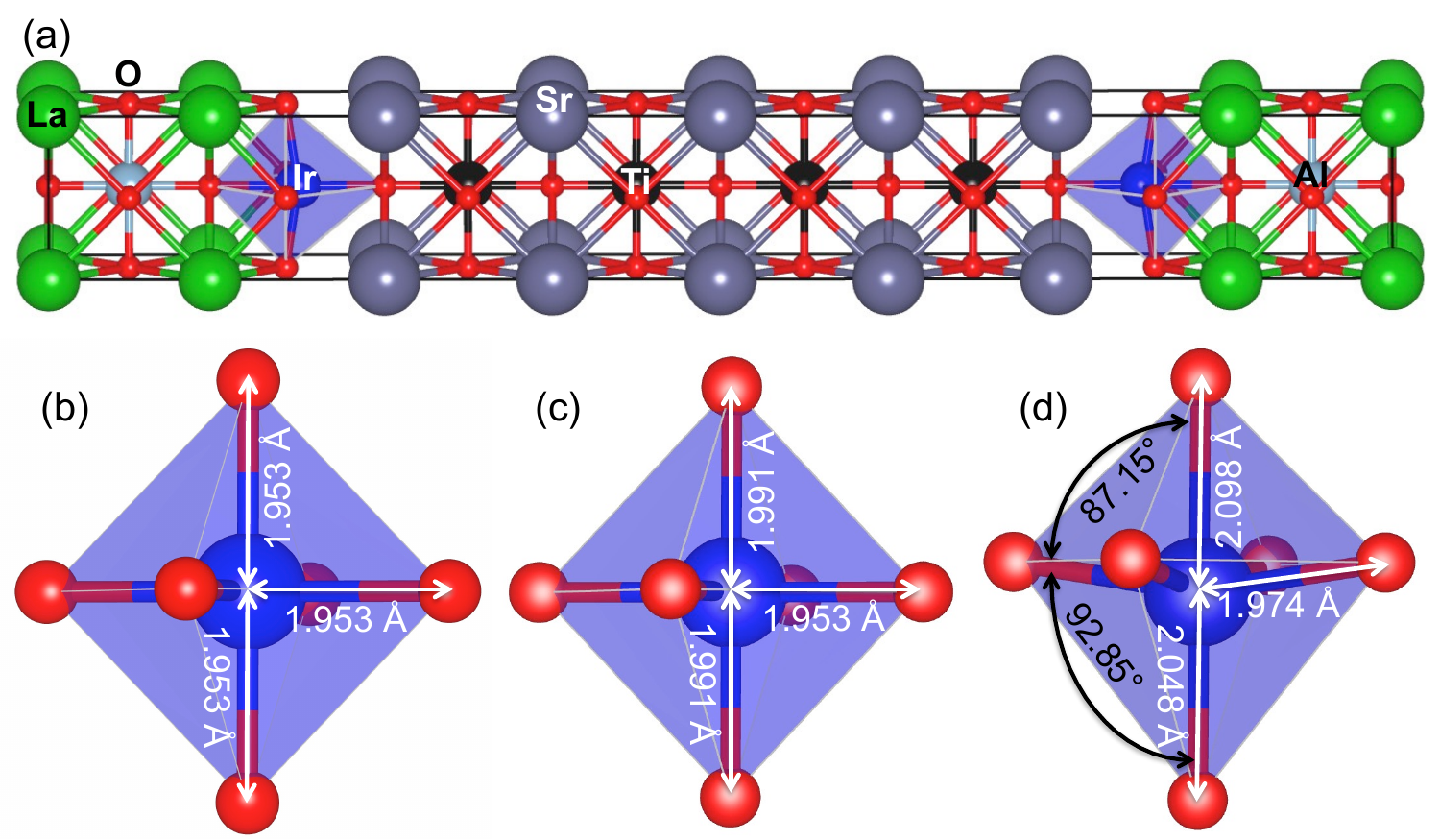}
    \caption{\label{fig:1a1bStructure}Panel (a) depicts a simulation cell of $1a \times 1b$ heterostructure of (LaAlO$_3$)$_{2.5}|$(SrIrO$_3$)$_1|$(SrTiO$_3$)$_{3.5}|$(SrIrO$_3$)$_1$ along the $c$-direction, allowing for the possibility of inversion symmetry, labeling each type of atom. The progressively distorted IrO$_6$ octahedra are illustrated for (a) the undistorted heterostructure, (b) the heterostructure with tetragonal JT-like distortion, and (c) the heterostructure with distorted bond lengths and bond angles, marking the corresponding bond lengths and bond angles.}
\end{figure}
We have simulated n-type interfaces in a superlattice unit of LAO$_{1.5}|$SIO$_1|$STO$_{3.5}|$SIO$_1|$LAO$_{1.5}$ along the $c$-direction within periodic boundary condition, as shown in \cref{fig:1a1bStructure}(a) to allow for the possibility of inversion symmetry and to ensure half an electron charge transfer to the interfaces, as discussed in Ref.~\cite{ChakrabortyPRB20, GanguliPRL14}. While in an epitaxial heterostructure, the amount of charge required to be transferred to the interface would be dictated by the electrostatic principles discussed in Ref.~\cite{GanguliPRL14}, our simulated structure ensures a transfer of 0.5 electrons per interface unit cell to satisfy the preferred oxidation states of various ions. Such an arrangement is consistent with the observation based on electrostatics that an n-type interface of LAO$|$SIO$|$STO heterostructure will rapidly reach the saturation transferred charge density of 0.5 electrons per interface unit cell \cite{ChakrabortyPRB20}.

The work has been carried out using {\em ab initio} calculations based on density functional theory (DFT) and analytical model calculations, with the latter being described in detail in the \cref{sec:RDmodels}. The DFT calculations for the LaAlO$_3|$SrIrO$_3|$SrTiO$_3$ heterostructures considered here were performed using the projector augmented wave (PAW) method \cite{paw} in conjunction with a plane-wave basis set, as implemented in the {\scshape vasp} code \cite{vasp1,vasp2}. A generalized gradient approximation (GGA) due to \citet{pbe} along with a Hubbard-$U$ correction for the transition metal $d$-orbitals due to \citet{DudarevPRB98}, henceforth referred to as the GGA+$U$ method, was invoked for approximating the exchange-correlation functional, with a moderate value of $U - J = 1.5$, 3, and 10~eV for Ir-$5d$, Ti-$3d$, and La-$4f$ orbitals, respectively \cite{ChakrabortyPRB20}. The Hubbard $U$ correction used for different elements serve different purposes. The empty La-$4f$ states place themselves at a much lower energy than expected. We use a large $U$ value for these states to correct their placement in energy \cite{GanguliPRL14}. The relatively small $U$ value for the Ti-$3d$ states helps correct the description of any transferred electron to those states \cite{GanguliPRL14}. Ir-$5d$ states host a weak magnetism in the $2a \times 2b$ heterostructure when described within a small Hubbard-$U$ correction \cite{ChakrabortyPRB20}. In order to make the results comparable with our previous works, we use the same $U$ values in our calculations for the $1a \times 1b$ heterostructure. Some key results were tallied against local density approximation (LDA) \cite{ldaCA, PerdewPRB81} + Hubbard $U$ method, hereafter referred to as the LDA+$U$ method. Unless mentioned otherwise for some specific cases, the lattice vectors and the atomic positions were optimized to minimize the stress and the Hellman-Feynman force on each atom to a tolerance of $10^{-2}$~eV/\AA, respectively, subject to a fixed lattice constant of 3.905~\AA\ in the $ab$-plane to match that of a thick STO substrate in practice. $\Gamma$-centered $15 \times 15 \times 1$ and $7 \times 7 \times 1$ $k$-point meshes were used for the integrations over the Brillouin zone using improved tetrahedron method \cite{BlochlPRB94T} for the calculations of three-dimensional (3D) band dispersions, isoenergetic contours and spin textures with $1a \times 1b$ and $2a \times 2b$ cells, respectively. Calculations accounting for spin-orbit interaction were performed with an energy convergence threshold of $10^{-5}$~eV.

\section{\label{sec:results}Results and Discussions}
We systematically discuss the electronic structure and spin-splitting in $1a \times 1b$ and $2a \times 2b$ heterostructures.

\subsection{$1a \times 1b$ heterostructure}
Since the present work focuses on the connection between the structural distortions and the corresponding Rashba-like spin-orbit interaction, carefully describing the structural distortions in different cases is the key to further progress. As the spin-orbit interaction arises primarily from the Ir ions, we carefully look into the local geometry of IrO$_6$ octahedra. The undistorted heterostructure with an ideal cubic perovskite form is considered with a lattice constant of 3.905~\AA, matching that of STO in all three directions. \cref{fig:1a1bStructure}(b) illustrates an IrO$_6$ octahedron belonging to this structure with all Ir-O bonds of equal length: 1.953~\AA. Considering tetragonal JT-like distortions in the structure, we note the distance between the Ir and the epical O ions to have increased to 1.991~\AA, while the Ir-O bond length in the $ab$-plane remained unchanged, as illustrated in \cref{fig:1a1bStructure}(c). The heterostructure with optimized atomic positions reveals that the O$^{2-}$ ions from the (IrO$_2$)$^0$ plane have moved closer to the (LaO)$^+$ plane due to electrostatic attraction (see \cref{fig:1a1bStructure}(a)), making the O-Ir-O bond angles 87.15$^\circ$ and 92.85$^\circ$, as seen from \cref{fig:1a1bStructure}(d). The bond lengths of Ir with the epical O atoms change to 2.048, 2.098~\AA, while the one between Ir and planer O atom changes to 1.974~\AA, as depicted in \cref{fig:1a1bStructure}(d).

With the idea of mapping different possible structural distortions in oxide heterostructures to bulk or structure inversion asymmetry and identifying the nature of the corresponding Rashba-like spin-orbit interaction, we consider the LAO$|$SIO$|$STO heterostructure with different levels of distortions: (a) the undistorted heterostructure, (b) the heterostructure with JT-like tetragonal distortions, and (c) the structure with optimized $\vec{c}$ lattice vector and optimally distorted bond lengths and bond angles. The undistorted structure has been designed from the space group $Pm\bar{3}m$ by filling the $1a$, $1b$, and $3c$ sites with Sr/La, Ti/Ir/Al, and O atoms, respectively. The space group preserves inversion symmetry, and the atomic sites $1a$, $1b$, and $3c$ correspond to centrosymmetric point groups $O_h$, $O_h$, and $D_{4h}$, respectively, leading to an overall $D_{4h}$ point symmetry. Further, the construction of the heterostructure with 3.5 unit cells of SrTiO$_3$ at the center, followed by one unit cell of SrIrO$_3$ on both sides, capped with 2.5 unit cells of LaAlO$_3$ distributed on both sides ensures inversion symmetry in the heterostructure with respect to the center. Hence, the possibility of any Rashba or Dresselhaus-type spin splitting is precluded due to the preserved bulk and structure inversion symmetry. The structure with JT-like tetragonal distortions is also expected to protect the overall inversion symmetry and the centrosymmetry at the point group level, as there are no significant structural changes, except systematic elongation of the bonds along the $\vec{c}$-direction. The structure with distorted bond lengths and angles is expected to belong to a tetragonal space group $P4mm$ with site symmetry point groups $C_{4v}$ and $C_{2v}$. However, the electrostatic forces near the interface attracted the O$^{2-}$ ions from the IrO$_2$ planes towards the (LaO)$^+$ planes (see \cref{fig:1a1bStructure}(a)), leading to further lowering of symmetry. We find an overall $C_1$ point symmetry for the system, lacking inversion symmetry. Next, we systematically analyze the electronic structure of the system in these structural forms to look for the implications of the progressive lowering of symmetry.

\begin{figure*}
	\includegraphics[scale = 1.15]{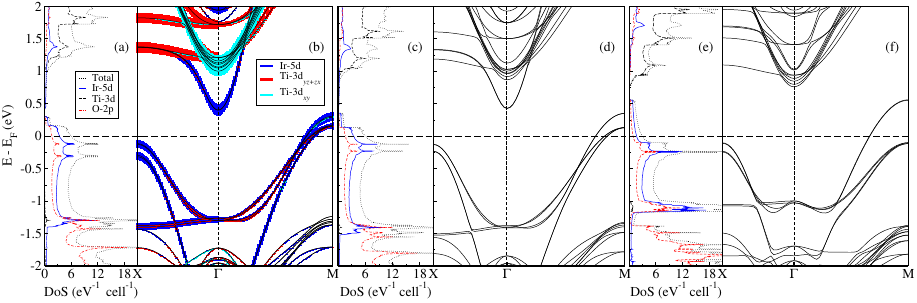}
    \caption{\label{fig:UnpolarizedBandDoS}The spin-unpolarized DoS, including orbital-projected DoS and band structure highlighting orbital characters along $X \to \Gamma \to M$ direction for $1a \times 1b$ LAO$|$SIO$|$STO undistorted heterostructure are shown in (a) and (b), respectively. Panels (c) and (d) represent similar DoS and band plots for the heterostructure accommodating tetragonal JT-like distortions. Panels (e) and (f) show the DoS and bands for the heterostructure with optimally distorted bond lengths and angles.}
\end{figure*}
At first, we examine the spin-unpolarized density of states (DoS) and band structures of the heterostructure, shown in \cref{fig:UnpolarizedBandDoS}, for the three different structural forms mentioned above. The DoS and the projected DoS in \cref{fig:UnpolarizedBandDoS}(a) reveal the conducting nature of the heterostructure with the states near the Fermi level comprising Ir-$5d$ orbitals hybridized with O-$2p$ orbitals. The Ti-$3d$ orbitals are found above the Fermi level, with the $3d_{yz}$ and $3d_{zx}$ orbitals higher in energy than the $3d_{xy}$ orbitals, as seen from the band dispersion in \cref{fig:UnpolarizedBandDoS}(b) \cite{GanguliPRL14}. While the $t_{2g}$ orbitals are degenerate in the bulk perovskites, confinement along the $\vec{c}$-direction in the heterostructure elevates the $d_{yz}$ and $d_{zx}$ bands in energy. From the electronic structure, we deduce that the nominal oxidation states for Ti and Ir are $4+$ and $3.5+$, respectively. The structural form accommodating JT-like tetragonal distortion reveals no significant change in the electronics structure (see \cref{fig:UnpolarizedBandDoS}(c), \cref{fig:UnpolarizedBandDoS}(d)), except the $d_{yz}$ and $d_{zx}$ bands coming closer to the $d_{xy}$ bands, owing to the elongation of bonds along the $\vec{c}$-direction, thereby reducing the effect of the confinement. The structure with optimally distorted bond lengths and bond angles exhibits some significant changes in the electronic structure, as seen from the DoS and the band dispersion in \cref{fig:UnpolarizedBandDoS}(e) and \cref{fig:UnpolarizedBandDoS}(f), respectively. We find the Ti-$3d$ and the Ir-$5d$ bands overlapping in energy above the Fermi level. The energy difference between the $d_{xy}$ bands and the $d_{yz}$, $d_{zx}$ bands further reduce upon accommodating the distortions, and some of the otherwise degenerate bands near the $\Gamma$-point become non-degenerate due to reduced symmetry (see \cref{fig:UnpolarizedBandDoS}(f)).
\begin{figure*}
	\includegraphics[scale = 1.15]{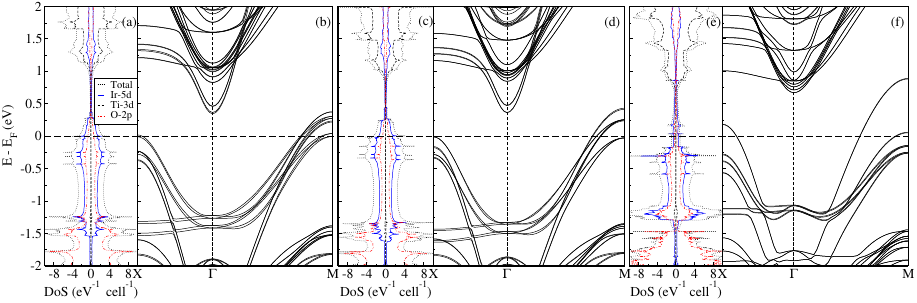}
    \caption{\label{fig:AFMnoSOCbandDoS}The spin-polarized DoS and band structure along $X \to \Gamma \to M$ direction for $1a \times 1b$ LAO$|$SIO$|$STO undistorted heterostructure are shown in (a) and (b), respectively. Panels (c) and (d) represent similar DoS and band plots for the heterostructure accommodating tetragonal JT-like distortions. Panels (e) and (f) show the DoS and bands for the heterostructure with distorted bond lengths and angles.}
\end{figure*}
Similar features are observed in the density of states and band dispersion upon spin-polarization (see \cref{fig:AFMnoSOCbandDoS}), except the small gap between the predominantly Ir-$5d$ and predominantly Ti-$3d$ bands seen in \cref{fig:UnpolarizedBandDoS} vanishes upon spin polarization. A small projected magnetic moment of 0.045~$\mu_B$ is observed at the Ir sites, while the Ti sites reveal no projected magnetic moment. The mirror-symmetric feature in the DoS corresponding to the two spins (see \cref{fig:AFMnoSOCbandDoS}(a), (c), (e)) indicates that the magnetic moments corresponding to the Ir atoms from the two different interfaces align opposite to each other. However, their significant separation makes hardly any exchange interaction possible between them. The band dispersions are shown in \cref{fig:AFMnoSOCbandDoS}(b), \cref{fig:AFMnoSOCbandDoS}(d), and \cref{fig:AFMnoSOCbandDoS}(f) corresponds to only one spin, without any loss of information, as the bands corresponding to the other spin exactly overlaps on them.

\begin{figure*}
	\includegraphics[scale = 1.15]{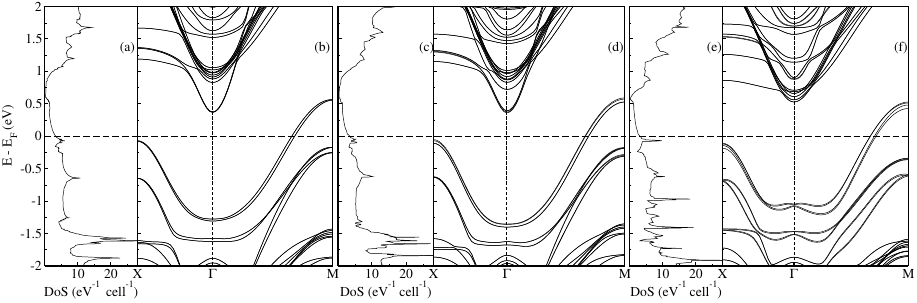}
	\caption{\label{fig:AFMsocBandDoS}The DoS and band structure, including spin-orbit interaction, along $X \to \Gamma \to M$ direction for $1a \times 1b$ LAO$|$SIO$|$STO undistorted heterostructure are shown in (a) and (b), respectively. Panels (c) and (d) represent similar DoS and band plots for the heterostructure accommodating tetragonal JT-like distortions. Panels (e) and (f) show the DoS and bands for the heterostructure with distorted bond lengths and angles.}
\end{figure*}
Upon introducing spin-orbit interaction, the density of states shown in \cref{fig:AFMsocBandDoS}(a), \cref{fig:AFMsocBandDoS}(c), and \cref{fig:AFMsocBandDoS}(e) for the undistorted structure, JT-like distorted structure, and the structure with distorted bond lengths and bond angles, with the corresponding band dispersions shown in \cref{fig:AFMsocBandDoS}(b), \cref{fig:AFMsocBandDoS}(d), and \cref{fig:AFMsocBandDoS}(f), respectively, do not show any significant change, except the Rashba-like splittings in the bands for the distorted structure (see \cref{fig:AFMsocBandDoS}(f)). A comparison with the projected DoS in \cref{fig:UnpolarizedBandDoS}(e) and \cref{fig:AFMnoSOCbandDoS}(e) reveals that the bands crossing the Fermi level in \cref{fig:AFMsocBandDoS}(f) primarily correspond to Ir-$5d$ orbitals, making them of particular interest in the context of Rashba-like interaction. We closely analyze these interesting bands by plotting the corresponding 3D bands, isoenergetic contours, and spin textures and comparing them with their counterparts from other heterostructures considered here.

\begin{figure*}
	\includegraphics[scale = 0.26]{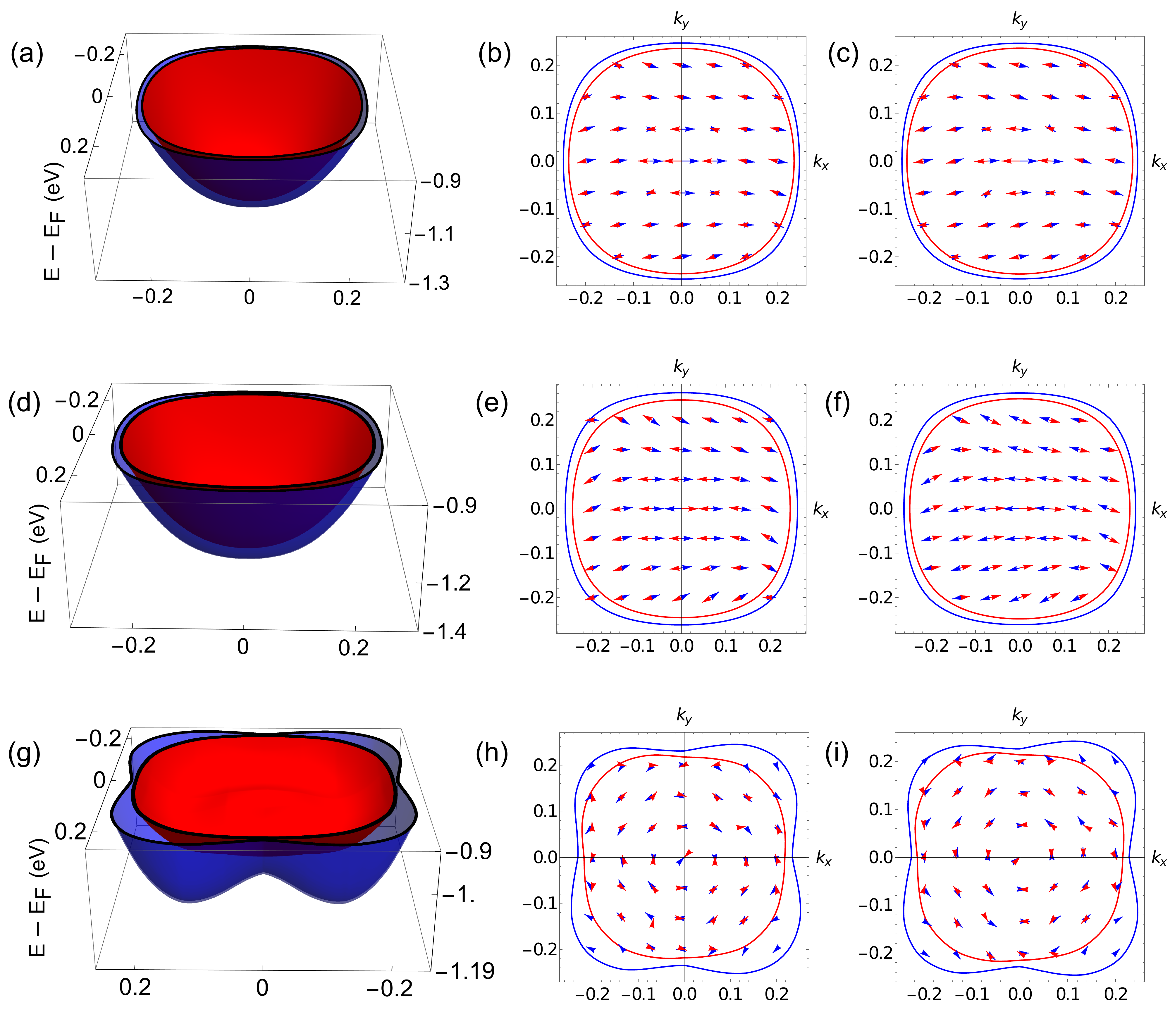}
	\caption{\label{fig:118SpinTexture}The 3D bands, the corresponding isoenergetic contours, and spin textures arising from two different SrIrO$_3$ layers are displayed in this figure. Panel (a) shows the 3D bands crossing the Fermi level for the undistorted heterostructure. In contrast, panel (b) (panel (c)) shows the isoenergetic contours at $E - E_F = -0.9$~eV and spin textures arising from two different SrIrO$_3$ layers. Similarly, panels (d), (e), and (f) show the 3D bands, isoenergetic contours at $E - E_F = -0.9$~eV, and spin textures for the heterostructure accommodating tetragonal JT-like distortions, while panels (g), (h), and (i) show the 3D bands, isoenergetic contours at $E - E_F = -0.9$~eV, and spin textures for the heterostructure with distorted bond lengths and bond angles.}
\end{figure*}
In order to understand the Rashba-like spin-orbit interaction in the system with progressive asymmetry, we plot the energy $\varepsilon$ as functions of $k_x$ and $k_y$ (so-called 3D bands) corresponding to the bands crossing the Fermi energy for the heterostructures with all three types of distortions, and the corresponding isoenergetic contours and spin textures in \cref{fig:118SpinTexture}. \Cref{fig:118SpinTexture}(a) displays the $\varepsilon(k_x, k_y)$ plot corresponding to the bands crossing the Fermi level for the undistorted heterostructure. \Cref{fig:118SpinTexture}(b) and \cref{fig:118SpinTexture}(c) show the corresponding isoenergetic contours and the projected spin directions marked using arrows arising from two different SrIrO$_3$ layers. All the 3D bands and the isoenergetic contours exhibited in \cref{fig:118SpinTexture} show a four-fold rotational symmetry, typical of $d_{yz}$ and $d_{zx}$ bands in the $k_x-k_y$ plane \cite{KumarPRB22}. The projected spin vectors $\vec{S}(\vec{k})$ at a crystal momentum $\vec{k} \equiv k_x \hat{x} + k_y \hat{y}$ are almost collinear and do not show a helical pattern in this case. The essential characteristic of a spin texture corresponding to Rashba-like spin-orbit interaction, {\em viz.} $\vec{S}(-\vec{k}) = -\vec{S}(\vec{k})$ for a particular band is also not observed here. Instead, the projected spins corresponding to a specific band exhibit $\vec{S}(-\vec{k}) = \vec{S}(\vec{k})$ behavior. Therefore, we conclude that the antiferromagnetic arrangement of spins leads to the opposite spins from different bands in the spin texture, while no Rashba-like interaction is observed in the undistorted heterostructure due to the symmetry being preserved. Owing to preserved overall inversion symmetry and centrosymmetry at the point group level, the heterostructure with JT-like tetragonal distortions also does not exhibit any Rashba-like interaction, as confirmed from the $\varepsilon(k_x, k_y)$ dispersion plot, the isoenergetic contours, and the spin textures shown in \cref{fig:118SpinTexture}(d),(e),(f). The Rashba-like spin helix obeying the condition $\vec{S}(-\vec{k}) = -\vec{S}(\vec{k})$ for a particular band is not observed here, although the spin vectors far away from the Brillouin zone center are not always collinear with those near the zone center.

In the case of the heterostructure with distorted bond lengths and bond angles, we find from the 3D bands in \cref{fig:118SpinTexture}(g), and the isoenergetic contours in \cref{fig:118SpinTexture}(h) and \cref{fig:118SpinTexture}(i) arising from two different SrIrO$_3$ layers that the bands crossing the Fermi level are almost degenerate at $k_x = 0$ and $k_y = 0$ lines, but have a large separation along the diagonals of the Brillouin zone. The projected spin textures corresponding to these bands, displayed in \cref{fig:118SpinTexture}(h) and \cref{fig:118SpinTexture}(i), reveal the spins orienting themselves tangentially to a circle, obeying the $\vec{S}(-\vec{k}) = -\vec{S}(\vec{k})$ feature for both layers of SrIrO$_3$. Such orientation of the spins confirms predominantly linear Rashba-like interaction in the system (see the appendix) owing to the compromised structure inversion symmetry due to the distortions and the microscopic electric field \cite{ChakrabortyPRB20}.

Our careful simulation of the heterostructure preserving inversion symmetry in the absence of distortion and the results with progressive distortions illustrate how broken inversion symmetry helps the Rashba effect set in.

\subsection{$2a \times 2b$ heterostructure}
\begin{figure}
	\includegraphics[scale = 0.54]{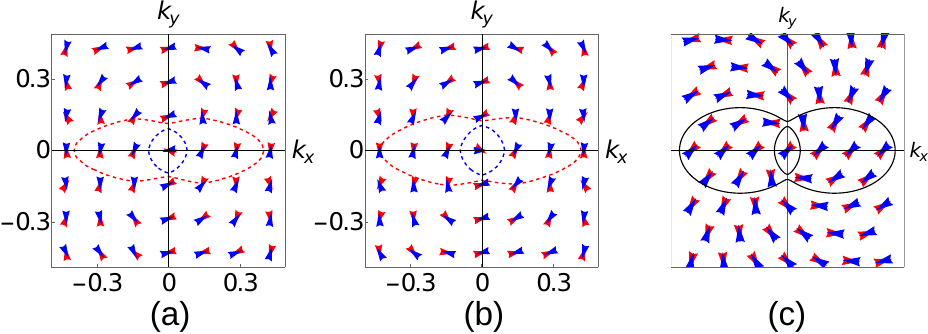}
	\caption{\label{fig:228SpinTexture}The isoenergetic contours at $E - E_F = 0.44$~eV and the spin textures directly obtained from DFT calculations for the $2a \times 2b$ structure are shown here. Panels (a) and (b) correspond to two different antiferromagnetic sublattices. Panel (c) shows the isoenergetic contours and spin textures obtained from a combined Rashba-Dresselhaus interaction model discussed in Ref.~\cite{ChakrabortyPRB20}.}
\end{figure}
After systematically studying the $1a \times 1b$ heterostructures with progressive distortions, we investigate the spin texture directly obtained from the DFT calculations in the $2a \times 2b$ heterostructure with tilted octahedra that we studied earlier in Ref.~\cite{ChakrabortyPRB20}. In our earlier work \cite{ChakrabortyPRB20}, we have shown that the magnetic moments in Ir atoms of a $2a \times 2b$ LAO$|$SIO$|$STO heterostructure align in a canted checkerboard antiferromagnetic fashion. Further, the bands near the Fermi level, having predominantly Ir-$5d$ character, show a combined Rashba-Dresselhaus splitting, agreeing with an analytical model calculation. We calculated the eigenstates $| \pm \rangle$ and the projected spin texture from the spin components $\langle \pm | S_x | \pm \rangle$, $\langle \pm | S_y | \pm \rangle$, and $\langle \pm | S_z | \pm \rangle$ using the same model \cite{ChakrabortyPRB20}. Here we calculate the projected spin texture directly from our DFT results, with the help of the spin components $\langle \psi_{\pm} | S_x | \psi_{\pm} \rangle$, $\langle \psi_{\pm} | S_y | \psi_{\pm} \rangle$, and $\langle \psi_{\pm} | S_z | \psi_{\pm} \rangle$, where $| \psi_{\pm} \rangle$ are the states corresponding to the relevant DFT bands. Our results, shown in \cref{fig:228SpinTexture}(a) and \cref{fig:228SpinTexture}(b) corresponding to two different antiferromagnetic sublattices, reveal the majority of the projected spins to be aligned along a diagonal line and flipping spin across the $k_x = 0$ line, exhibiting overall agreement with the model results in Ref.~\cite{ChakrabortyPRB20}, confirming the combined Rashba-Dresselhaus interaction model's validity. Although the bond lengths and angles were distorted in the $1a \times 1b$ structure, the tilted IrO$_6$ octahedra observed in the $2a \times 2b$ structure were not seen there. Such tilted octahedra introduce bulk inversion asymmetry in the system, in combination with structure inversion asymmetry, leading to a combined Rashba-Dresselhaus interaction.

\section{\label{sec:conc}Conclusion}
The role of inversion asymmetry in spin-orbit interaction-driven spin splitting, particularly bulk inversion asymmetry and structure inversion asymmetry leading to Dresselhaus and Rashba splitting, respectively, has earlier been theoretically established \cite{DresselhausPR55, RashbaSPSS60}. However, mapping the commonly occurring structural distortions in perovskite structures to bulk or structure inversion asymmetry is often challenging in perovskite oxide heterostructures due to their complex nature. Moreover, similar splitting patterns arising from different spin-orbit interactions result in further confusion. However, the proposed technological applications based on such spin splittings critically depend on the exact nature of the interaction, making understanding the nature of interaction imperative. Our study illustrates how inversion asymmetry in the presence of a microscopic electric field \cite{ChakrabortyPRB20} can lead to Rashba and Dresselhaus spin-orbit interaction using an example of a perovskite oxide heterostructure. Our first-principles calculations within density functional theory, 3D band dispersion, and a comparison of the spin textures obtained from our DFT calculations and theoretical models help us connect the structural distortions with the Rashba-like spin-orbit interaction pattern observed in such systems. Starting from a $1a \times 1b$ heterostructure of LaAlO$_3|$SrIrO$_3|$SrTiO$_3$ simulated in such a way that preserves the inversion symmetry, we allowed progressive distortions in the bond lengths and the bond angles, without the IrO$_6$ octahedra being tilted. Our results reveal linear Rashba spin-orbit interaction for the distorted $1a \times 1b$ heterostructure, suggesting only structure inversion asymmetry was realized through the distortions, while the bulk inversion symmetry was intact. However, considering a $2a \times 2b$ heterostructure allowed for tilted IrO$_6$ octahedra, leading to bulk and structure inversion asymmetry manifesting as a combined Rashba and Dresselhaus spin-orbit interaction. A comparison of the spin texture obtained from our DFT calculations for a $2a \times 2b$ heterostructure and a spin texture predicted from our previous model calculation are in excellent agreement \cite{ChakrabortyPRB20}, implying the model's validation. We have also considered various combinations of Rashba and Dresselhaus interactions, compiling an elaborate reference for the spin-splitting and spin textures corresponding to such interactions. Besides illustrating the connection between structural asymmetry and the Rashba-like spin-orbit interactions through an example of perovskite oxide heterostructures, our work helps classify via the spin textures similar spin-splitting due to spin-orbit interaction. Our results indicate that Jahn-Teller distortion or other distortions of bond length and bond angles in perovskite oxides do not lead to bulk inversion asymmetry, although it may be enough to break the overall inversion symmetry in a heterostructure. However, octahedral tilts observed in this perovskite heterostructure is a systematic structural distortion leading to bulk inversion asymmetry and Dresselhaus interaction in the $2a \times 2b$ heterostructure. Thus, we observe Rashba and Dresselhaus interactions coexisting in the $2a \times 2b$ heterostructure. Our techniques of analyzing the spin textures may help study topological materials and understand the corresponding transport properties.

\begin{acknowledgments}
Financial support from SERB, India, through grants number CRG/2021/005320, number ECR/2016/001004, INSPIRE fellowship from DST, India through grant number IF171000, and the use of high-performance computing facilities of IISER Bhopal are gratefully acknowledged.
\end{acknowledgments}

\appendix* \section{\label{sec:RDmodels}A manual for identifying the nature of Rashba-like interactions}
The spin-orbit interaction of the Rashba or Dresselhaus type arises due to broken inversion symmetry in the structure. The interaction Hamiltonian consists of products of the crystal momentum components and Pauli matrices that remain invariant under all the symmetry operations of the system, leading to spin-split bands and a helical spin texture. Therefore, one must first find the underlying symmetry and point group to find an appropriate spin Hamiltonian for a given system. Since spin splitting happens due to inversion asymmetry, the odd terms in $k$, particularly the linear and the cubic terms, contribute the most. However, due to their structural complexity, finding the underlying symmetry may often be difficult in distorted perovskite oxide compounds, heterostructures, and superlattices. Moreover, the structural symmetry often does not predict the relative strength between the linear and the cubic terms. To avoid such confusion, here we collect some linear and cubic Rashba and Dresselhaus interaction terms and show the corresponding band dispersions, isoenergetic contours, and spin textures for the purpose of identifying the nature of the interaction in a given system.

Various Rashba-like spin-orbit interactions of different orders and their combinations may be identified by comparing the conventional band dispersions, 3D band dispersions, isoenergetic contours, and projected spin textures obtained from theoretical or experimental techniques with that obtained from the relevant model(s) \cite{ChakrabortyPRB20, KumarPRB22}. Below, we write a representative model Hamiltonian for Rashba-Dresselhaus interactions in the first and third order and evaluate the eigenvalues and eigenstates corresponding to some useful combinations. Since we consider a representative model Hamiltonian for illustration purposes, all the terms included in the model may not remain invariant under a given symmetry point group.

\subsection{The model}
A Rashba-Dresselhaus model Hamiltonian combining all terms relevant to the present context may be written, assuming an electric field along the $z$-direction, as
\begin{equation}
	H = H_0 + H_{R_{(1)}} + H_{R_{(3)}} + H_{D_{(1)}} + H_{D_{(3)}}, \label{eq:Hamiltonian}
\end{equation}
where $H_0$ represents the free-electron Hamiltonian in two dimensions:
\begin{equation}
	H_0 = - \frac{1}{2m^*} \left( \frac{\partial^2}{\partial x^2} + \frac{\partial^2}{\partial y^2} \right),
\end{equation}
with $m^*$ denoting the effective mass, $H_{R_{(1)}}$ and $H_{R_{(3)}}$ represent the linear and cubic Rashba interactions, respectively,
\begin{align}
	H_{R_{(1)}} = & \alpha_{(1)} i (k_- \sigma_+ - k_+ \sigma_-) = \alpha_{(1)} (k_y \sigma_x - k_x \sigma_y), \\
	H_{R_{(3)}} = & \alpha_{(3)} i (k_-^3 \sigma_+ - k_+^3 \sigma_-) \nonumber \\
	= & \alpha_{(3)} [(3k_x^2 - k_y^2)k_y\sigma_x - (k_x^2 - 3k_y^2)k_x\sigma_y],
\end{align}
and $H_{D_{(1)}}$ and $H_{D_{(3)}}$ represent the linear and cubic Dresselhaus interactions, respectively,
\begin{align}
	H_{D_{(1)}} = & -\beta_{(1)} (k_+ \sigma_+ + k_- \sigma_-) = \beta_{(1)} (k_y \sigma_y - k_x \sigma_x), \\
	H_{D_{(3)}} = & - \beta_{(3)} (k_+^2 - k_-^2)(k_- \sigma_+ - k_+ \sigma_-) \nonumber \\
	= & - 4 \beta_{(3)} k_x k_y (k_y \sigma_x - k_x \sigma_y),
\end{align}
with
\begin{align}
	k_+ &= k_x + ik_y; & k_x &= \frac{1}{2}(k_+ + k_-); \nonumber \\
	k_- &= k_x - ik_y; & k_y &= \frac{1}{2i}(k_+ - k_-); \nonumber \\
	\sigma_+ &= \frac{1}{2}(\sigma_x + i \sigma_y); & \sigma_x &= (\sigma_+ + \sigma_-); \nonumber \\
	\sigma_- &= \frac{1}{2}(\sigma_x - i \sigma_y); & \sigma_y &= \frac{1}{i}(\sigma_+ - \sigma_-), \label{eq:relations}
\end{align}
where $k_x, k_y$ represent the components of the reciprocal space vector, $\sigma_x, \sigma_y$ represent the Pauli matrices, and $\alpha_{(1)}$, $\alpha_{(3)}$ ($\beta_{(1)}$, $\beta_{(3)}$) represent the linear and cubic Rashba (Dresselhaus) coefficients, respectively. Next, we solve the models in appropriate combinations for eigenvalues and eigenstates and evaluate the expectation values of the spin component operators to find the corresponding helical spin texture.
\subsection{Linear Rashba interaction}
\begin{figure*}
	\includegraphics[scale = 0.56]{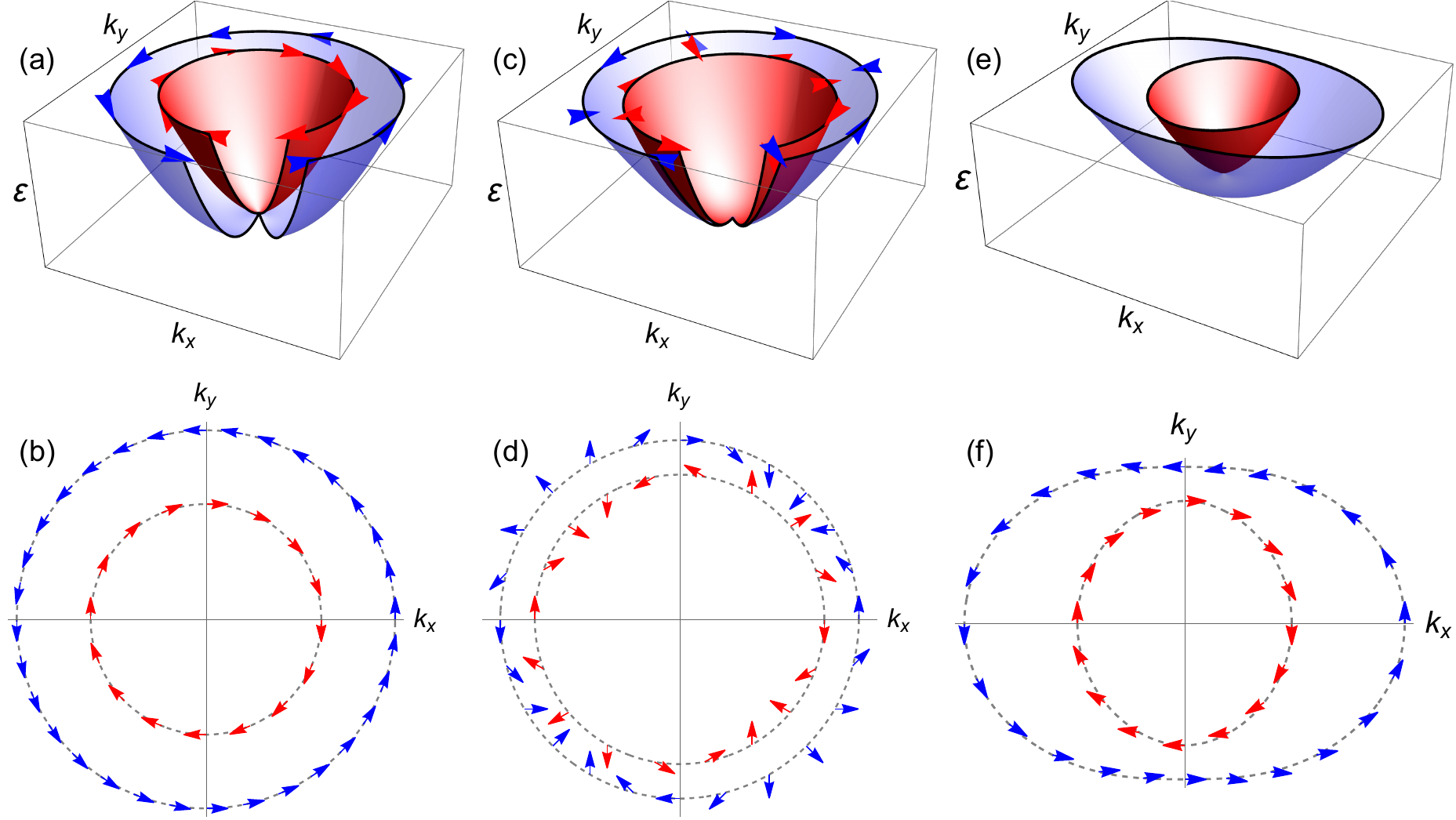}
	\caption{\label{fig:RashbaModel}The band structures, isoenergetic contours, and the projected spin directions corresponding to the Rashba interactions of different orders have been illustrated here. Panels (a) and (b) depict the 3D band dispersion $\varepsilon_{R_{(1)}}^{\pm}(\vec{k})$ and isoenergetic contours in the $k_x$-$k_y$ plane, respectively, marked with the projected spin directions for the linear Rashba interaction. Similarly, panels (c) and (d) illustrate the 3D band dispersion $\varepsilon_{R_{(3)}}^{\pm}(\vec{k})$ and the isoenergetic contours, respectively, marked with the projected spin directions for the cubic Rashba interaction; panels (e) and (f) illustrate the 3D bands $\varepsilon_{R_{(1,3)}}^{\pm}(\vec{k})$ and isoenergetic contours marked with projected spin directions, respectively, for a combined Rashba interaction of linear and cubic order.}
\end{figure*}
First, we consider only the linear Rashba interaction by omitting the $H_{R_{(3)}}$, $H_{D_{(1)}}$, and $H_{D_{(3)}}$ terms in \cref{eq:Hamiltonian}. The energy eigenvalues for the Hamiltonian may be obtained as
\begin{equation}
	\varepsilon_{R_{(1)}}^{\pm}(\vec{k}) = \frac{k^2}{2m^*} \pm \alpha_{(1)}k, ~~~\text{where } k \equiv |\vec{k}|, \label{eq:R1EigenValues}
\end{equation}
as illustrated in \cref{fig:RashbaModel}(a), with eigenstates
\begin{equation}
	|\pm\rangle_{R_{(1)}} = \frac{1}{\sqrt{2}} \left[ |\uparrow\rangle \mp i \exp(i \phi) |\downarrow\rangle \right], \label{eq:R1EigenStates}
\end{equation}
$\phi$ representing the polar angle in $k$-space satisfying $k_x = k \cos \phi$, $k_y = k \sin \phi$, and $(|\uparrow\rangle, |\downarrow\rangle)$ representing the eigenstates of the spin projection operator $S_z$. Once the eigenstates are obtained, the projected spin components along $x$, $y$, and $z$ may be evaluated as follows:
\begin{align}
	\langle S_x \rangle_{R_{(1)}}^+ &= \frac{1}{2} \langle + | \sigma_x | + \rangle_{R_{(1)}} = - \frac{i}{4}[\exp(i \phi) - \exp(-i \phi)] \nonumber \\
	&= \frac{1}{2} \sin \phi = - \frac{1}{2} \langle - | \sigma_x | - \rangle_{R_{(1)}} = - \langle S_x \rangle_{R_{(1)}}^-, \nonumber \\
	\langle S_y \rangle_{R_{(1)}}^+ &= \frac{1}{2} \langle + | \sigma_y | + \rangle_{R_{(1)}} = \frac{i}{4}[i \exp(i \phi) + i \exp(-i \phi)] \nonumber \\
	&= - \frac{1}{2} \cos \phi = - \frac{1}{2} \langle - | \sigma_y | - \rangle_{R_{(1)}} = - \langle S_y \rangle_{R_{(1)}}^-, \nonumber \\
	\langle S_z \rangle_{R_{(1)}}^+ &= \frac{1}{2} \langle + | \sigma_z | + \rangle_{R_{(1)}} = \frac{1}{4} \{1 + i^2 \exp[i(\phi - \phi)]\} \nonumber \\
	&= 0 = \frac{1}{2} \langle - | \sigma_z | - \rangle_{R_{(1)}} = \langle S_z \rangle_{R_{(1)}}^-. \label{eq:R1spin}
\end{align}
The isoenergetic contours in the 2D $k$-space may be deduced by setting the energy eigenvalues in \cref{eq:R1EigenValues} to a constant $\varepsilon_c$ and solving for the roots of the quadratic equations:
\begin{align}
	\varepsilon_c &= \frac{k^2}{2m^*} \pm \alpha_{(1)} k, ~~\text{implies} \nonumber \\
	k^{\pm} &= m^* \left( \mp \alpha_{(1)} \pm \sqrt{\alpha_{(1)}^2 + \frac{2 \varepsilon_c}{m^*}} \right), \label{eq:R1Isoenergetic}
\end{align}
where only two roots are physically acceptable. Since $k^{\pm}$ in \cref{eq:R1Isoenergetic} turn out to be independent of $\phi$, the isoenergetic contours would be two concentric circles centered at the origin of the 2D $k$-space, as shown in \cref{fig:RashbaModel}(a) and \cref{fig:RashbaModel}(b), marked with the projected spin directions deduced in \cref{eq:R1spin}.
\subsection{Cubic Rashba interaction}
Similar to the above, here we consider only cubic Rashba interaction, omitting the $H_{R_{(1)}}$, $H_{D_{(1)}}$, and $H_{D_{(3)}}$ terms in \cref{eq:Hamiltonian}. The eigenvalues and eigenstates for this model turn out to be
\begin{align}
	\varepsilon_{R_{(3)}}^{\pm} (\vec{k}) &= \frac{k^2}{2m^*} \pm \alpha_{(3)} k^3, \text{ and } \label{eq:R3Eigenvalues} \\
	| \pm \rangle_{R_{(3)}} &= \frac{1}{\sqrt{2}} [| \uparrow \rangle \mp i \exp(3i \phi) | \downarrow \rangle ],
\end{align}
respectively. \cref{fig:RashbaModel}(c) depicts the 3D band structure obtained from \cref{eq:R3Eigenvalues}. The projected spin components for this case take the form:
\begin{align}
	\langle S_x \rangle_{R_{(3)}}^{\pm} &= \frac{1}{2} \langle \pm | \sigma_x | \pm \rangle_{R_{(3)}} = \pm \frac{1}{2} \sin 3\phi, \nonumber \\
	\langle S_y \rangle_{R_{(3)}}^{\pm} &=  \frac{1}{2} \langle \pm | \sigma_y | \pm \rangle_{R_{(3)}} = \mp \frac{1}{2} \cos 3\phi, \nonumber \\
	\langle S_z \rangle_{R_{(3)}}^{\pm} &= \frac{1}{2} \langle \pm | \sigma_z | \pm \rangle_{R_{(3)}} = 0. \label{eq:R3spin}
\end{align}
The isoenergetic contours for these bands in the $k_x$-$k_y$ plane may be given by the physically acceptable roots of the cubic equation
\begin{equation}
	\varepsilon_c = \frac{k^2}{2m^*} \pm \alpha_{(3)} k^3,
\end{equation}
which also trace concentric circles around the $\Gamma$-point, as seen in \cref{fig:RashbaModel}(d). However, the projected spin directions deduced in \cref{eq:R3spin} and illustrated in \cref{fig:RashbaModel}(d) for cubic Rashba interaction exhibit a precession frequency thrice that of linear Rashba interaction.
\subsection{Combined linear and cubic Rashba interactions}
Rashba interaction of mixed linear and cubic order may be described by omitting the terms $H_{D_{(1)}}$ and $H_{D_{(3)}}$ in \cref{eq:Hamiltonian}. The corresponding eigenvalues are given as
\begin{align}
	& \varepsilon_{R_{(1,3)}}^{\pm} (\vec{k}) \nonumber \\
	=& \frac{k^2}{2m^*} \pm \sqrt{\alpha_{(1)}^2k^2 + \alpha_{(3)}^2k^6 + 2 \alpha_{(1)} \alpha_{(3)} k^2(k_x^2 - k_y^2)} \nonumber \\
	=& \frac{k^2}{2m^*} \pm \sqrt{\alpha_{(1)}^2k^2 + \alpha_{(3)}^2k^6 + 2 \alpha_{(1)} \alpha_{(3)} k^4 \cos 2\phi}, \label{eq:R13Eigenvalues}
\end{align}
with eigenstates
\begin{align}
	|\pm \rangle_{R_{(1,3)}} &= \frac{1}{\sqrt{2}} (\pm \zeta_{R_{(1,3)}} |\uparrow \rangle + |\downarrow \rangle), \text{ where} \nonumber \\
	\zeta_{R_{(1,3)}} &= i \frac{\sqrt{\alpha_{(1)}^2 + \alpha_{(3)}^2 k^4 + 2 \alpha_{(1)} \alpha_{(3)} k^2 \cos 2\phi}}{\alpha_{(1)} \exp(i \phi) + \alpha_{(3)} k^2 \exp(3i \phi)}, \nonumber \\ \zeta_{R_{(1,3)}}^* \zeta_{R_{(1,3)}} &= 1.
\end{align}
The projected spin components for these eigenstates may be given as
\begin{align}
	\langle S_x \rangle_{R_{(1,3)}}^+ &= \frac{1}{2} \langle + | \sigma_x | + \rangle_{R_{(1,3)}} = \frac{1}{4} \left(\zeta_{R_{(1,3)}} + \zeta_{R_{(1,3)}}^* \right) \nonumber \\
	&= \frac{1}{2} \frac{\alpha_{(1)} \sin \phi + \alpha_{(3)} k^2 \sin 3\phi}{\sqrt{\alpha_{(1)}^2 + \alpha_{(3)}^2 k^4 + 2 \alpha_{(1)} \alpha_{(3)} k^2 \cos 2\phi}} \nonumber \\
	&= -\frac{1}{2} \langle - | \sigma_x | - \rangle_{R_{(1,3)}} = - \langle S_x \rangle_{R_{(1,3)}}^-, \nonumber \\
	\langle S_y \rangle_{R_{(1,3)}}^+ &= \frac{1}{2} \langle + | \sigma_y | + \rangle_{R_{(1,3)}} = \frac{i}{4} \left(\zeta_{R_{(1,3)}} - \zeta_{R_{(1,3)}}^* \right) \nonumber \\
	&= - \frac{1}{2} \frac{\alpha_{(1)} \cos \phi + \alpha_{(3)} k^2 \cos 3\phi}{\sqrt{\alpha_{(1)}^2 + \alpha_{(3)}^2 k^4 + 2 \alpha_{(1)} \alpha_{(3)} k^2 \cos 2\phi}} \nonumber \\
	&= -\frac{1}{2} \langle - | \sigma_y | - \rangle_{R_{(1,3)}} = - \langle S_y \rangle_{R_{(1,3)}}^-, \nonumber \\
	\langle S_z \rangle_{R_{(1,3)}}^+ &= \frac{1}{2} \langle + | \sigma_z | + \rangle_{R_{(1,3)}} = \frac{1}{4} \left(\zeta_{R_{(1,3)}}^* \zeta_{R_{(1,3)}} - 1 \right) = 0 \nonumber \\
	&= \frac{1}{2} \langle - | \sigma_z | - \rangle_{R_{(1,3)}} = \langle S_z \rangle_{R_{(1,3)}}^-. \label{eq:R13Spin}
\end{align}
The isoenergetic contours may be obtained by setting the energy eigenvalues in \cref{eq:R13Eigenvalues} to a constant and evaluating the physically acceptable roots. The 3D energy bands (eigenvalues) and the isoenergetic contours marked with the projected spin directions, as obtained from \cref{eq:R13Spin}, have been illustrated in \cref{fig:RashbaModel}(e) and \cref{fig:RashbaModel}(f), respectively. The outer and the inner isoenergetic contours (see \cref{fig:RashbaModel}(f)) roughly resemble ellipses with the major axes being perpendicular to each other.
\subsection{Linear Dresselhaus interaction}
\begin{figure*}
	\includegraphics[scale = 0.52]{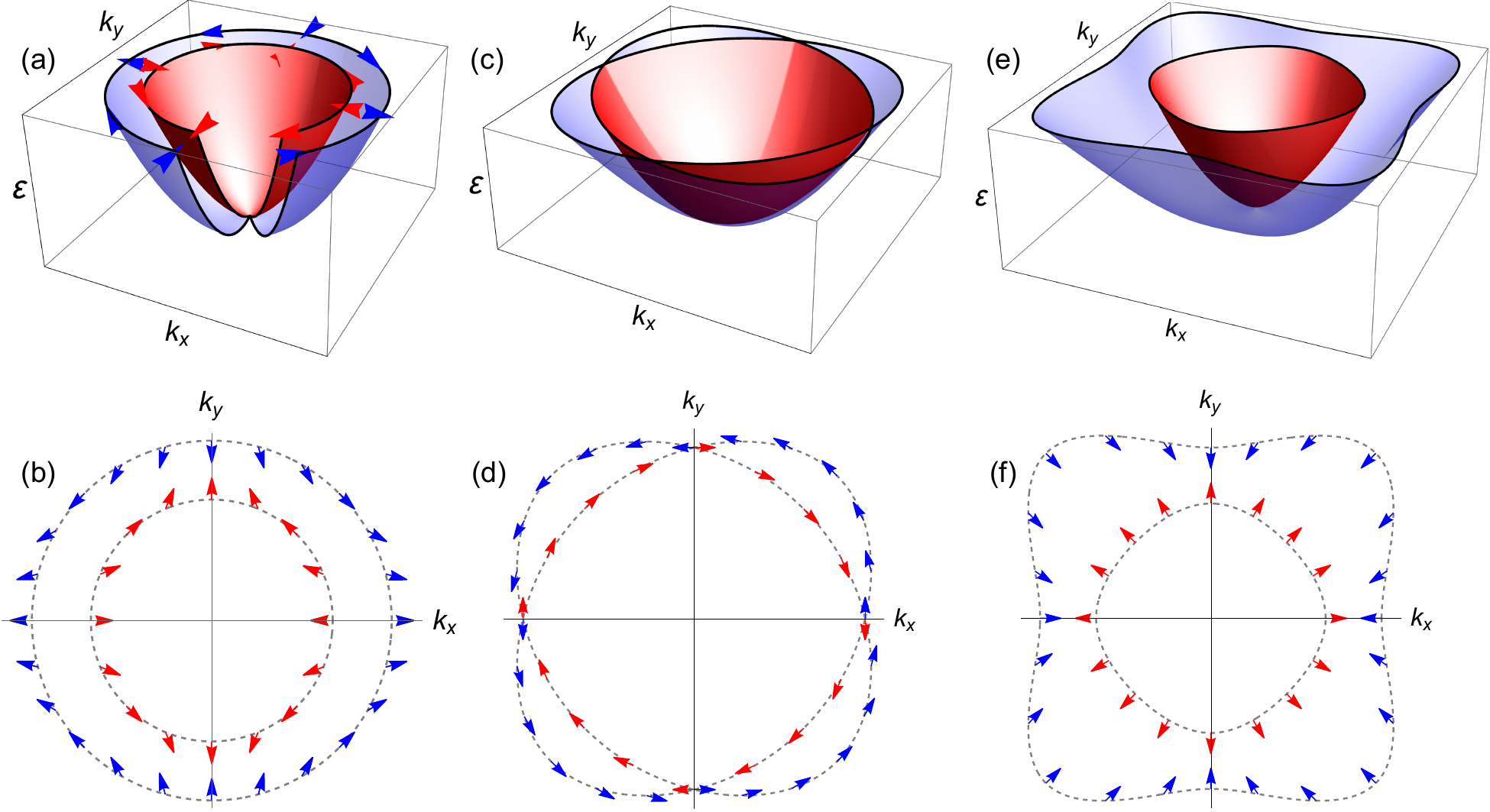}
	\caption{\label{fig:DresselhausModel}The band structures, isoenergetic contours, and the projected spin directions corresponding to the Dresselhaus interactions of different orders have been illustrated here. Panels (a) and (b) depict the 3D band dispersion $\varepsilon_{D_{(1)}}^{\pm}(\vec{k})$ and isoenergetic contours in the $k_x$-$k_y$ plane, respectively, marked with the projected spin directions for the linear Dresselhaus interaction. Similarly, panels (c) and (d) illustrate the 3D band dispersion $\varepsilon_{D_{(3)}}^{\pm}(\vec{k})$ and the isoenergetic contours marked with the projected spin directions respectively, for the cubic Dresselhaus interaction; panels (e) and (f) illustrate the 3D bands $\varepsilon_{D_{(1,3)}}^{\pm}(\vec{k})$ and isoenergetic contours marked with projected spin directions, respectively, for a combined Dresselhaus interaction of linear and cubic order.}
\end{figure*}
In order to understand the implications of the linear Dresselhaus interaction, we omit the $H_{R_{(1)}}$, $H_{R_{(3)}}$, and $H_{D_{(3)}}$ terms in \cref{eq:Hamiltonian}; then evaluate the energy eigenvalues and eigenstates as
\begin{equation}
	\varepsilon_{D_{(1)}}^{\pm} (\vec{k}) = \frac{k^2}{2m^*} \pm \beta_{(1)} k
\end{equation}
and
\begin{equation}
	| \pm \rangle_{D_{(1)}} = \frac{1}{\sqrt{2}} (|\uparrow\rangle \pm \exp(-i \phi) |\downarrow\rangle), \label{eq:D1EigenStates}
\end{equation}
respectively. While the eigenvalues are similar to that of linear Rashba eigenvalues (see \cref{eq:R1EigenValues}), the eigenstates differ in a nontrivial way from \cref{eq:R1EigenStates}. The projected spin components for this case may be evaluated as
\begin{align}
	\langle S_x \rangle_{D_{(1)}}^{\pm} &= \frac{1}{2} \langle \pm | \sigma_x | \pm \rangle_{D_{(1)}} = \mp \frac{1}{2} \cos \phi, \nonumber \\
	\langle S_y \rangle_{D_{(1)}}{\pm} &= \frac{1}{2} \langle \pm | \sigma_y | \pm \rangle_{D_{(1)}} = \pm \frac{1}{2} \sin \phi, \nonumber \\
	\langle S_z \rangle_{D_{(1)}}^{\pm} &= \frac{1}{2} \langle \pm | \sigma_z | \pm \rangle_{D_{(1)}} = 0. \label{eq:D1spin}
\end{align}
The similarity between the energy eigenvalues of linear Rashba and linear Dresselhaus Hamiltonians indicates that the isoenergetic contours in the case of linear Dresselhaus spin-orbit interaction would also trace concentric circles. The band dispersion, as seen in \cref{fig:DresselhausModel}(a), is similar to that of linear Rashba interaction, although the projected spin directions shown on the 3D bands in \cref{fig:DresselhausModel}(a) and on the isoenergetic contours in \cref{fig:DresselhausModel}(b) significantly differ. In contrast with linear Rashba interaction, where all spins were aligned along the tangent of the isoenergetic contours (see \cref{fig:RashbaModel}(b)), here the spins precess between tangential and normal orientations to the contours.
\subsection{Cubic Dresselhaus interaction}
We investigate the Dresselhaus interaction of cubic order by omitting the $H_{R_{(1)}}$, $H_{R_{(3)}}$, and $H_{D_{(1)}}$ terms in \cref{eq:Hamiltonian}. Solving for the energy eigenvalues, we get
\begin{equation}
	\varepsilon_{D_{(3)}}^{\pm} (\vec{k}) = \frac{k^2}{2m^*} \pm 2 \beta_{(3)} k^3 | \sin 2\phi |.
\end{equation}
Unlike the cubic Rashba energy eigenvalues (see \cref{eq:R3Eigenvalues} and \cref{fig:RashbaModel}(a),(b)), the eigenvalues in the present case lack circular symmetry, as seen from the 3D band dispersion and isoenergetic contours in \cref{fig:DresselhausModel}(c) and \cref{fig:DresselhausModel}(d), respectively. Instead, we find a four-fold rotational symmetry for the bands. The corresponding eigenstates are the same as that of linear Rashba interaction:
\begin{equation}
	|\pm\rangle_{D_{(3)}} = \frac{1}{\sqrt{2}} [|\uparrow\rangle \mp i \exp(i \phi) |\downarrow\rangle ],
\end{equation}
because the cubic Dresselhaus term $H_{D_{(3)}}$ commutes with the linear Rashba term $H_{R_{(1)}}$. One need not separately evaluate the projected spin components for this case since it would be the same as \cref{eq:R1spin}, shown in \cref{fig:DresselhausModel}(d) overlaid on the isoenergetic contours.
\subsection{Combined linear and cubic Dresselhaus interactions}
We study the combined linear and cubic Dresselhaus interaction by omitting the $H_{R_{(1)}}$ and $H_{R_{(3)}}$ terms in \cref{eq:Hamiltonian}. The corresponding energy eigenvalues,
\begin{align}
	\varepsilon_{D_{(1,3)}}^{\pm} (\vec{k}) =& \frac{k^2}{2m^*} \nonumber \\
	& \pm k \sqrt{\beta_{(1)}^2 + 4 \beta_{(3)} k^2 \sin^2 2\phi \left( \beta_{(1)} + \beta_{(3)} k^2 \right)},
\end{align}
are depicted in \cref{fig:DresselhausModel}(e). The eigenstates may be evaluated as
\begin{align}
	| \pm \rangle_{D_{(1,3)}} &= \frac{1}{\sqrt{2}} (\pm \zeta_{D_{(1,3)}} |\uparrow\rangle + |\downarrow\rangle), \text{ where} \nonumber \\
	\zeta_{D_{(1,3)}} &= \frac{\sqrt{\beta_{(1)}^2 + 4 \beta_{(3)} k^2 \left(\beta_{(1)} + \beta_{(3)} k^2 \right) \sin^2 2\phi}}{\beta_{(1)} \exp(-i \phi) - 2i \beta_{(3)} k^2 \sin 2\phi \exp(i \phi)}, \nonumber \\
	\zeta_{D_{(1,3)}}^* \zeta_{D_{(1,3)}} &= 1.
\end{align}
The projected spin components may be given as
\begin{align}
	\langle S_x \rangle_{D_{(1,3)}}^+ &= \frac{1}{2} \langle + | \sigma_x | + \rangle_{D_{(1,3)}} = \frac{1}{4}(\zeta_{D_{(1,3)}} + \zeta_{D_{(1,3)}}^*) \nonumber \\
	&= \frac{1}{2} \frac{\beta_{(1)} \cos \phi + 2 \beta_{(3)} k^2 \sin \phi \sin 2\phi}{\sqrt{\beta_{(1)}^2 + 4 \beta_{(3)} k^2 \sin^2 2\phi \left(\beta_{(1)} + \beta_{(3)} k^2 \right)}} \nonumber \\
	&= -\frac{1}{2} \langle - | \sigma_x | - \rangle_{D_{(1,3)}} = -\langle S_x \rangle_{D_{(1,3)}}^-, \nonumber \\
	\langle S_y \rangle_{D_{(1,3)}}^+ &= \frac{1}{2} \langle + | \sigma_y | + \rangle_{D_{(1,3)}} = \frac{i}{4}(\zeta_{D_{(1,3)}} - \zeta_{D_{(1,3)}}^*) \nonumber \\
	&= \frac{1}{2} \frac{\beta_{(1)} \sin \phi + 2 \beta_{(3)} k^2 \cos \phi \sin 2\phi}{\sqrt{\beta_{(1)}^2 + 4 \beta_{(3)} k^2 \sin^2 2\phi \left(\beta_{(1)} + \beta_{(3)} k^2 \right)}} \nonumber \\
	&= -\frac{1}{2} \langle - | \sigma_y | - \rangle_{D_{(1,3)}} = -\langle S_y \rangle_{D_{(1,3)}}^-, \nonumber \\
	\langle S_z \rangle_{D_{(1,3)}}^+ &= \frac{1}{2} \langle + | \sigma_z | + \rangle_{D_{(1,3)}} = \left(\zeta_{D_{(1,3)}}^* \zeta_{D_{(1,3)}} - 1 \right) \nonumber \\
	&= 0 = \frac{1}{2} \langle - | \sigma_z | - \rangle_{D_{(1,3)}} = \langle S_z \rangle_{D_{(1,3)}}^-.
\end{align}
Overlaying the projected spin directions on the isoenergetic contours (see \cref{fig:DresselhausModel}(f)) reveals that the spins prefer to stay nearly normal to the contours, with some exceptions on the outer contour.
\subsection{Combined linear Rashba and linear Dresselhaus interactions}
\begin{figure}
	\centering
	\includegraphics[scale = 0.43]{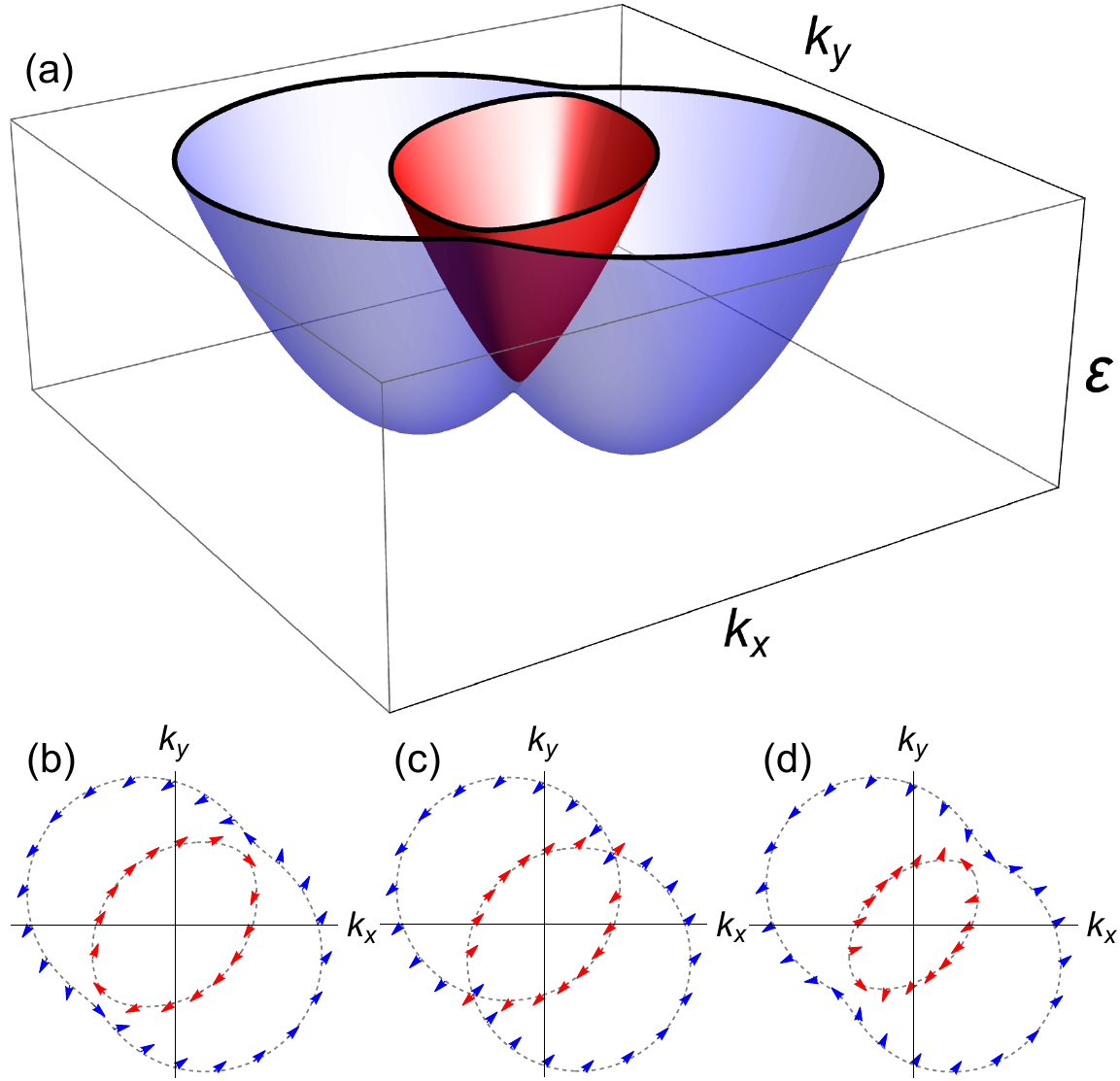}
	\caption{\label{fig:R1D1Model}The 3D band dispersions $\varepsilon_{R_{(1)}D_{(1)}}^{\pm}(\vec{k})$ for a combination of linear Rashba and linear Dresselhaus interactions are illustrated in (a). The isoenergetic contours marked with the projected spin directions are depicted for (b) $\alpha_{(1)} > \beta_{(1)}$, (c) $\alpha_{(1)} = \beta_{(1)}$, and (d) $\alpha_{(1)} < \beta_{(1)}$.}
\end{figure}
Strong spin-orbit interaction combined with bulk and structure inversion asymmetry and an electric field leads to a combined Rashba-Dresselhaus interaction, where the $k$-linear terms dominate over any term of $k^3$ order. We obtain the relevant Hamiltonian by omitting the $k$-cubic terms $H_{R_{(3)}}$ and $H_{D_{(3)}}$ in \cref{eq:Hamiltonian}. The corresponding energy eigenvalues may be obtained as \cite{ChakrabortyPRB20}
\begin{align}
	\varepsilon_{R_{(1)}D_{(1)}}^{\pm} (\vec{k}) &= \frac{k^2}{2m^*} \pm \sqrt{\left(\alpha_{(1)}^2 + \beta_{(1)}^2 \right)k^2 - 4 \alpha_{(1)} \beta_{(1)} k_x k_y} \nonumber \\
	&= \frac{k^2}{2m^*} \pm k \sqrt{\alpha_{(1)}^2 + \beta_{(1)}^2 - 2 \alpha_{(1)} \beta_{(1)} \sin 2\phi},
\end{align}
as illustrated in \cref{fig:R1D1Model}(a), with eigenstates
\begin{align}
	| \pm \rangle &= \frac{1}{\sqrt{2}}(\pm \zeta_{R_{(1)}D_{(1)}} |\uparrow\rangle + |\downarrow\rangle), \text{ where} \nonumber \\
	\zeta_{R_{(1)}D_{(1)}} &= i \frac{\sqrt{\alpha_{(1)}^2 + \beta_{(1)}^2 - 2 \alpha_{(1)} \beta_{(1)} \sin 2\phi}}{\alpha_{(1)} \exp(i \phi) - i \beta_{(1)} \exp(-i \phi)}, \nonumber \\
	\zeta_{R_{(1)}D_{(1)}}^* \zeta_{R_{(1)}D_{(1)}} &= 1.
\end{align}
The projected spin components may be computed as
\begin{align}
	\langle S_x \rangle_{R_{(1)}D_{(1)}}^+ &= \frac{1}{2} \langle + | \sigma_x | + \rangle_{R_{(1)}D_{(1)}} \nonumber \\
	&= \frac{1}{4} \left(\zeta_{R_{(1)}D_{(1)}} + \zeta_{R_{(1)}D_{(1)}}^* \right) \nonumber \\
	&= \frac{1}{2} \frac{\alpha_{(1)} \sin \phi - \beta_{(1)} \cos \phi}{\sqrt{\alpha_{(1)}^2 + \beta_{(1)}^2 - 2 \alpha_{(1)} \beta_{(1)} \sin 2\phi}} \nonumber \\
	&= -\frac{1}{2} \langle - | \sigma_x | - \rangle_{R_{(1)}D_{(1)}} = -\langle S_x \rangle_{R_{(1)}D_{(1)}}^-, \nonumber \\
	\langle S_y \rangle_{R_{(1)}D_{(1)}}^+ &= \frac{1}{2} \langle + | \sigma_y | + \rangle_{R_{(1)}D_{(1)}} \nonumber \\
	&= \frac{1}{4} \left(\zeta_{R_{(1)}D_{(1)}} - \zeta_{R_{(1)}D_{(1)}}^* \right) \nonumber \\
	&= \frac{1}{2} \frac{\beta_{(1)} \sin \phi - \alpha_{(1)} \cos \phi}{\sqrt{\alpha_{(1)}^2 + \beta_{(1)}^2 - 2 \alpha_{(1)} \beta_{(1)} \sin 2\phi}} \nonumber \\
	&= -\frac{1}{2} \langle - | \sigma_y | - \rangle_{R_{(1)}D_{(1)}} = -\langle S_y \rangle_{R_{(1)}D_{(1)}}^-, \nonumber \\
	\langle S_z \rangle_{R_{(1)}D_{(1)}}^+ &= \frac{1}{2} \langle + | \sigma_z | + \rangle_{R_{(1)}D_{(1)}} \nonumber \\
	&= \frac{1}{4} \left(\zeta_{R_{(1)}D_{(1)}}^* \zeta_{R_{(1)}D_{(1)}} - 1 \right) \nonumber \\
	&= 0 = \frac{1}{2} \langle - | \sigma_z | - \rangle_{R_{(1)}D_{(1)}} = \langle S_z \rangle_{R_{(1)}D_{(1)}}^-.
\end{align}
While the 3D band structure reveals a double-minima pattern (see \cref{fig:R1D1Model}(a)), the isoenergetic contours shown in \cref{fig:R1D1Model}(b), \cref{fig:R1D1Model}(c), and \cref{fig:R1D1Model}(d) for $\alpha_{(1)} > \beta_{(1)}$, $\alpha_{(1)} = \beta_{(1)}$, and $\alpha_{(1)} < \beta_{(1)}$, respectively, indicate roughly two circles overlapping onto each other. Equal values of Rashba and Dresselhaus coefficients ($\alpha_1 = \beta_1$, see \cref{fig:R1D1Model}(c)) result in two perfect circles overlapping to form the contour, with the spins arranged only in parallel or anti-parallel fashion, leading to the so-called persistent spin helix \cite{BernevigPRL06, KoralekN09}.
\subsection{Combining all terms}
\begin{figure*}
	\includegraphics[scale = 0.67]{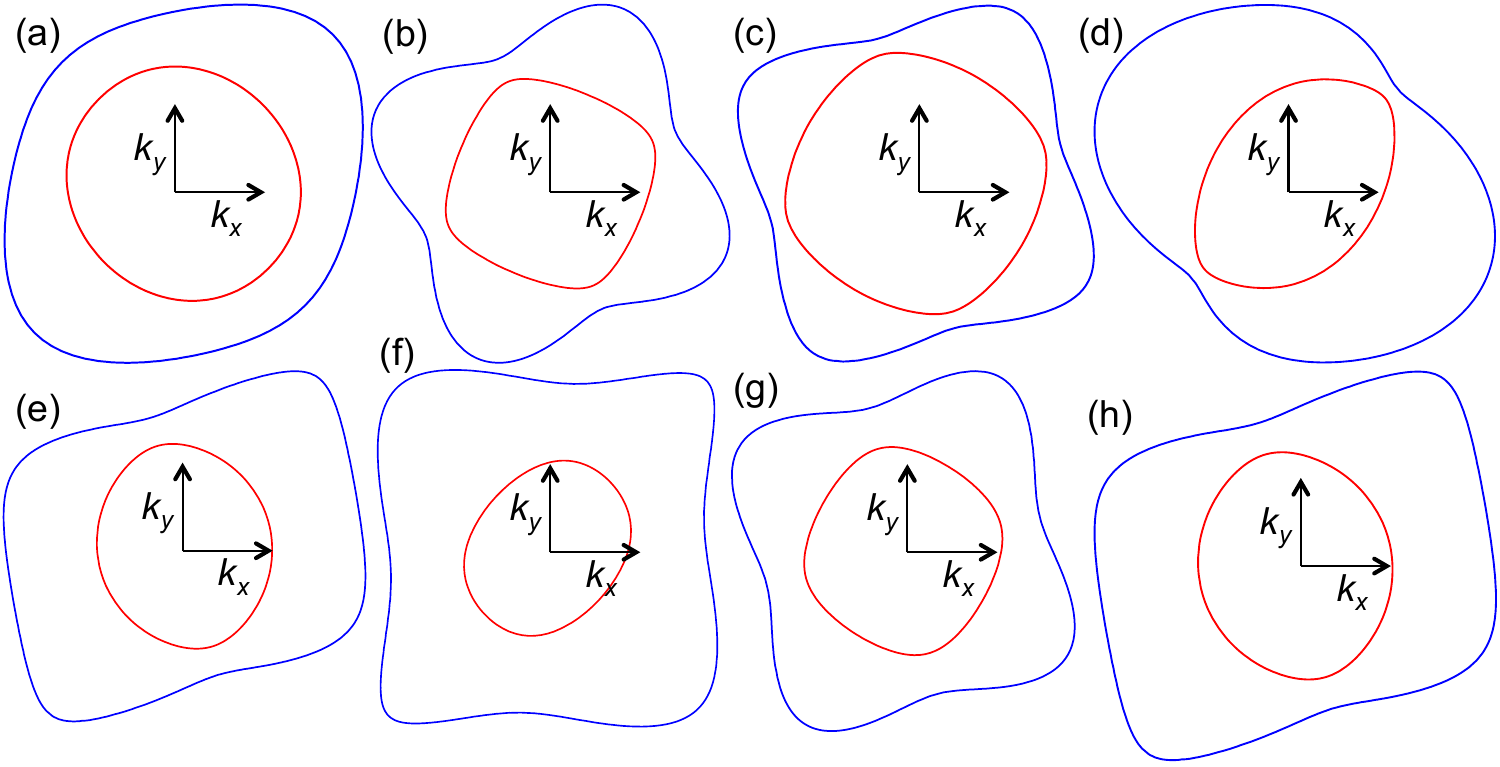}
	\caption{\label{fig:RemainingModel}The isoenergetic contours obtained from the eigenvalues of the Hamiltonian including all Rashba-Dresselhaus terms and setting different coefficients to zero as required are illustrated here. The panels represent (a) a combination of linear Rashba and cubic Dresselhaus interactions, (b) a combination of cubic Rashba and linear Dresselhaus interactions, (c) a combination of cubic Rashba and cubic Dresselhaus interactions, (d) a combination of linear Rashba, cubic Rashba, and linear Dresselhaus interactions, (e) a combination of linear Rashba, cubic Rashba, and cubic Dresselhaus interactions, (f) a combination of linear Dresselhaus, cubic Dresselhaus, and linear Rashba interactions, (g) a combination of linear Dresselhaus, cubic Dresselhaus, and cubic Rashba interactions, (h) a combination of all four interactions.}
\end{figure*}
In order to understand the pattern of the other possible combinations of Rashba and Dresselhaus interactions not discussed yet, we consider the full Hamiltonian in \cref{eq:Hamiltonian} and evaluate the eigenvalues as
\begin{align}
	\varepsilon_\text{full}^{\pm}(\vec{k}) =& \frac{k^2}{2m^*} \pm k [ \alpha_{(1)}^2 + \beta_{(1)}^2 + 2 \beta_{(1)} \beta_{(3)} k^2 \nonumber \\
	&+ (\alpha_{(3)}^2 + 2 \beta_{(3)}^2) k^4 + 2 \alpha_{(1)} \alpha_{(3)} k^2 \cos 2\phi \nonumber \\
	&- 2 \beta_{(3)} k^2 (\beta_{(1)} + \beta_{(3)} k^2) \cos 4\phi \nonumber \\
	&- 2\alpha_{(1)} \beta_{(1)} \sin 2\phi + 2 \alpha_{(1)} \beta_{(3)} k^2 \sin 2\phi  \nonumber \\
	&- 2 \alpha_{(3)} \beta_{(1)} k^2 \sin 4\phi - 2 \alpha_{(3)} \beta_{(3)} k^4 \sin 4\phi ]^{1/2}. \label{eq:GeneralEigenvalue}
\end{align}
One can obtain the eigenvalues of all possible combinations of Rashba and Dresselhaus terms given in \cref{eq:Hamiltonian} by setting different coefficients in \cref{eq:GeneralEigenvalue} to zero, as required. Here, we plot the isoenergetic contours for various combinations of Rashba-Dresselhaus terms to understand the geometry of the bands, as displayed in \cref{fig:RemainingModel}. A comparison of \Cref{fig:RashbaModel,fig:DresselhausModel,fig:R1D1Model,fig:RemainingModel} reveal that while individual linear Rashba (see \cref{fig:RashbaModel}(b)), cubic Rashba (see \cref{fig:RashbaModel}(d)), and linear Dresselhaus (see \cref{fig:DresselhausModel}(b)) interactions exhibit a circular symmetry of bands in an isoenergetic plane, cubic Dresselhaus interaction leads to a four-fold rotational symmetry (see \cref{fig:DresselhausModel}(d)). Further, combinations of linear Dresselhaus and cubic Dresselhaus interactions (see \cref{fig:DresselhausModel}(f)), cubic Rashba and linear Dresselhaus interactions (see \cref{fig:RemainingModel}(b)), cubic Rashba and cubic Dresselhaus interactions (see \cref{fig:RemainingModel}(c)), and cubic Rashba, linear Dresselhaus, and cubic Dresselhaus interactions (see \cref{fig:RemainingModel}(g)) also exhibit four-fold rotational symmetry of the isoenergetic contours, with the latter three being somewhat twisted in $k_x$-$k_y$-plane, reflecting the impact of the terms containing $\sin 4\phi$ in \cref{eq:GeneralEigenvalue}. All other combined interactions reveal only a two-fold rotational symmetry of bands in an isoenergetic plane.
\subsection{Anisotropic effective mass}
The effective mass characterizing the underlying free-electron-like system depends on the structure of the material and its symmetries. Obviously, the effective mass along different directions may be different for a structure with certain asymmetries. Some of the common distortions in oxide heterostructures may lead to anisotropy in effective mass, where the effective mass components along $x$ and $y$ directions are unequal: $m_x^* \neq m_y^*$ \cite{ChakrabortyPRB20}. Here we discuss the simple cases of linear Rashba and linear Dresselhaus interactions subject to anisotropic effective mass, as described by the Hamiltonians
\begin{align}
	H_{AR_{(1)}} &= H_{A0} + \alpha_{(1)} (k_y \sigma_x - k_x \sigma_y) \text{ and} \\
	H_{AD_{(1)}} &= H_{A0} + \beta_{(1)} (k_y \sigma_y - k_x \sigma_x),
\end{align}
respectively, with
\begin{equation}
	H_{A0} = -\frac{1}{2m_x^*} \frac{\partial^2}{\partial x^2} -\frac{1}{2m_y^*} \frac{\partial^2}{\partial y^2}.
\end{equation}
The corresponding energy eigenvalues may be expressed as
\begin{equation}
	\varepsilon_{AR_{(1)}(D_{(1)})}^{\pm}(\vec{k}) = \left( \frac{\cos^2 \phi}{2m_x^*} + \frac{\sin^2 \phi}{2m_y^*} \right) k^2 \pm \alpha_{(1)} (\beta_{(1)}) k. \label{eq:AnisoRashbaEigenvalues}
\end{equation}
The anisotropy in the effective mass would lead to elliptical isoenergetic contours with a common major axis. However, the eigenstates and the corresponding projected spin texture would remain unaltered.


%
\end{document}